\documentclass[aps,pre,showpacs,superscriptaddress,twocolumn,floatfix]{revtex4}

\usepackage{graphicx,psfrag}
\usepackage{dcolumn}
\usepackage{amssymb}
\usepackage{epsf,epsfig}

\begin{document}

\title{Guiding-fields for phase-separation: Controlling
Liesegang patterns}

\author{T. Antal}
\affiliation{Program for Evolutionary Dynamics, 
Harvard University, Cambridge, MA 02138}
\author{I. Bena}
\affiliation{Theoretical Physics Department, 
University of Geneva, CH-1211 Geneva 4, Switzerland}
\author{M. Droz}
\affiliation{Theoretical Physics Department, 
University of Geneva, CH-1211 Geneva 4, Switzerland}
\author{K. Martens}
\affiliation{Theoretical Physics Department, 
University of Geneva, CH-1211 Geneva 4, Switzerland}
\author{Z. R\'acz}
\affiliation{Institute for Theoretical Physics--HAS, 
E\"otv\"os University, P\'azm\'any s\'et\'any 1/a, 1117 Budapest, Hungary}

\date{\today}

\begin{abstract}
Liesegang patterns emerge from precipitation processes
and may be used to build bulk structures at submicron
lengthscales. Thus they have
significant potential for technological applications
provided adequate methods of control can be devised.
Here we describe a simple, physically realizable
pattern-control based on the notion
of {\it driven precipitation} meaning that the phase-separation
is governed by a {\it guiding field}
such as, for example, a temperature or a {\it pH} field.
The phase-separation is modeled through a non-autonomous
Cahn-Hilliard equation whose spinodal is determined
by the evolving guiding field. Control over the
dynamics of the spinodal gives control over
the velocity of the instability front which separates the stable
and unstable regions of the system. Since the
wavelength of the pattern is largely determined by this velocity,
the distance between successive
precipitation bands becomes controllable. We demonstrate the above
ideas by numerical studies of a
$1D$ system with diffusive guiding field.
We find that the results can be accurately described by
employing a linear stability analysis (pulled-front theory)
for determining the velocity -- local-wavelength relationship.
From the perspective of the Liesegang theory, our results
indicate that the so-called revert patterns may be naturally
generated by diffusive guiding fields.

\end{abstract}


\maketitle

\section{Introduction}
\label{introduction}

Pattern formation is an ubiquitous phenomenon in out-of-equilibrium
systems, and  ordered structures often emerge in the wake of a
moving reaction front~\cite{cross94}. There has been recently increasing 
interest, both experimental and theoretical, in the study of various types of 
chemically-generated patterns. The main reason is that they are expected
to provide new 
{\it bottom-up}, self-assembling  technologies for engineering bulk patterns
on mesoscopic and microscopic scales (for illustration see
e.g.~\cite{grzy3,grzy2,grzy1,optical,tsapatis,giraldo}
from a rapidly-growing bibliography), for which the traditional
{\it top-down}
methods (i.e., removing material in order to create a structure) 
are reaching the limits of their capabilities.

Detailed understanding of the mechanisms responsible
for pattern formation
is a key-element in developing technological
applications since it helps constructing the appropriate
tools for the control of the characteristics of the emerging patterns.
In this paper we shall focus on designing a simple method to  control 
the so-called {\em Liesegang structures}~\cite{liesegang1896,henisch91}. In
particular, this method should be useful for a recently-proposed
experimental set-up that allows to create {\it stamps}
of such structures~\cite{grzy1}. 

Depending on the geometry, Liesegang precipitation patterns
are bands (in an axially-symmetric configuration), rings or
shells (in a circular, respectively spherical-symmetric configuration),
clearly separated in the direction of motion of a chemical reaction
front. Several generic experimental laws characterize the patterns,
see e.g.~\cite{racz99,michel00} for reviews.
In particular, it is found  
that the positions of the bands usually obey simple laws; e.g.
they form, with a good approximation, a geometric series 
with increasing distance
between consecutive bands. This is the so-called 
{\it regular banding} situation,
which has been recently explained~\cite{antal99} using the 
{\it phase separation in the presence of a moving front} as the 
underlying mechanism. Briefly, the reaction front, which moves diffusively,
leaves behind a constant concentration $c_0$ of the {\it reaction product},
that we shall conventionally name hereafter  
{\it $C$ particles}~\cite{Note-on-C,antal98,droz99}.
At a coarse-grained level, the
dynamics of the $C$ particles 
(that can diffuse, and are also attracting 
each other) can be described by a Cahn-Hilliard 
equation~\cite{cahn58,cahn61,hohenberg} with a source term 
corresponding to the moving reaction front.
Starting with a system free of $C$'s, the dynamics of the front
brings locally the
system across the spinodal line, provided that $c_0$ is inside the
unstable region of the phase diagram. 
A phase separation takes then place on a
short time scale and a band of precipitate is rapidly formed just
behind the front. This band acts as a sink for the $C$ particles. Then the
local concentration of $C$'s decreases, bringing the system locally in
the stable phase again. Thus Liesegang patterns are formed since
the state of the system at the front is locally and
quasi-periodically driven into the unstable regime.

The characteristics of these {\em regular patterns} can be controlled
to some extent through an
appropriate choice of the concentration of the reagents~\cite{matalon},
of the nature of the gel that is filling the reaction 
container~\cite{toramaru}, the shape
of the container~\cite{optical}, or through an applied  electric
field~\cite{lagzi,sultan,shreif,bena}.

The spinodal decomposition scenario has proved its
power by describing {\em regular} patterns and, furthermore,
by explaining how those patterns can be influenced by the
concentration of the outer and inner electrolytes and by an
external electric field. We will show that it can
be extended to describe other situations, as well. Indeed,
there is experimental evidence of Liesegang-type 
precipitation patterns with decreasing  distances between
successive bands~\cite{inverse} which is termed {\it inverse banding}.
In the {\it borderline} case between regular and inverse banding,
the distances between successive bands are constant, 
a situation called {\it equidistant banding}~\cite{ortoleva}.
In our attempts of describing the above patterns we were lead
to a mechanism which may provide a
simple, experimentally realizable {\em control tool}
of the emerging pattern.

As described in detail in Sec.~\ref{model}, our proposal is based on a
phase separation mechanism in a {\em space- and time-dependent 
guiding field}, which could represent, for example, a temperature or a
{\em pH} field. The pattern formation is thus modeled through a 
{\em non-autonomous CH equation},  whose spinodal line is
controlled by the guiding field.
Note that the present design of the guiding field is different from the
{\it homogeneous (overall) cooling} that was used in most of the
previous studies of non-autonomous CH models,
see e.g.~\cite{vollmer,huston,alt,rebelo}.
As we shall demonstrate,
a simple guiding field is sufficient to generate crossover
between regular and inverse patterns.
For example, such a guiding field can be a
temperature field evolving diffusively due to a temperature 
difference at the boundaries 
of the system, whose characteristics are detailed in Sec.~\ref{gfc}.
The features of the corresponding
emerging patterns are analyzed in Sec.~\ref{results}.
As discussed in Sec.~\ref{interpretation}, our numerical findings can be
justified by theoretical arguments relating the velocity of the
front of the guiding field to the pulled-front velocity resulting
from a linear
stability analysis of the phase-separation process. Other, more flexible 
ways to control the patterns are also briefly presented in
Sec.~\ref{controling}. 
Finally,  conclusions and perspectives are
discussed in Sec.~\ref{conclusions}.

\section{The model} 
\label{model}

Let us consider a tube filled with gel,
and an initially uniform concentration $c_0$
of $C$ particles throughout the tube. 
Assuming axial symmetry along the $x$-axis of the tube,
we shall consider that the $C$-particle concentration
$c(x,t)$ evolves in time according to the  
Cahn-Hilliard (CH) equation in one dimension. 
After rescaling the space and time variables, this equation 
can be written in the following dimensionless form
\begin{equation}
\frac{\partial c(x,t)}{\partial t} = -\frac{\partial^2}{\partial x^2} 
\left[ \varepsilon c(x,t) - c^3(x,t) +  
\frac{\partial^2 c(x,t)}{\partial x^2}\right]\,, 
\label{ch1}
\end{equation}
with $0\leqslant x\leqslant L$, where 
$L$ is the dimensionless length of the tube.
Note that we also performed an appropriate {\em shift and scaling of
the concentration}  that allows us to write the CH equation
in a form that is more convenient for the exposition of our problem;
namely, this form is symmetric with respect to the change in sign
of $c$, $c\leftrightarrow -c$ (see e.g.~\cite{antal99,racz99}
and footnote~\cite{foot2} 
for a more detailed discussion of this point). The shifted and
rescaled concentration can take both positive and negative values,
and the stable configurations are symmetric around $c=0$. 

The parameter $\varepsilon$ measures the deviation of the 
temperature from the critical temperature $T_c$; it is negative for 
temperatures above $T_c$ (for which no phase-separation is possible),
while it is positive for temperatures $T<T_c$. Below the critical
temperature, a uniform concentration profile $c_0$
inside the spinodal decomposition domain, 
i.e.  $|c_0| \leqslant c_s=\sqrt{\varepsilon /3}$, is linearly unstable.
A small, localized perturbation of the concentration can then trigger 
a phase separation  throughout the system, through the amplification 
of the unstable modes of wavenumbers $|k|<\sqrt{\varepsilon}$.
A large body of work (see e.g.~\cite{bray})
has been devoted to the study of the phase separation process
in the simple case of a uniform parameter
$\varepsilon$  taking the same
value throughout the system.

Here we shall concentrate on a different situation, namely
when $\varepsilon$
is a {\em field}, which evolves according to its own dynamics.
Moreover, it is possible to control its evolution. For a simple
realization of this control consider the following situation.
Suppose that at a time $t=0$ the temperature at one end of the
tube is lowered and kept at a constant value $T_0<T_c$ thereafter.
The other end of the tube is supposed to be
thermally isolated~\cite{Tube-length}.
The temperature profile then evolves in time along the tube
according to the usual Fourier law of heat conduction, 
and so does the related $\varepsilon(x,t)$
field, 
\begin{equation}
\frac{\partial \varepsilon (x,t)}{\partial t} = D\, \frac{\partial^2\,  
\varepsilon (x,t)}{\partial x^2}\,, 
\label{ch2}
\end{equation}
where $D$ is the dimensionless thermal diffusion coefficient.
Through an appropriate scaling
of the temperature,  the value of $\varepsilon$ can be set to $-1$ 
throughout the system at $t=0$, while $\varepsilon=+1$ at $t\geqslant 0$ at the left 
($x=0$) end of the tube; at the right ($x=L$) 
end of the tube there is no heat flow:
\begin{eqnarray}
&&\varepsilon(x>0,t=0)=-1\,,\nonumber\\
&&\varepsilon(x=0,t)=+1\,,\nonumber\\
&&\frac{\partial \varepsilon}{\partial x}(x=L,t)=0\,.
\label{bic2}
\end{eqnarray}
These equations~(\ref{ch2}), (\ref{bic2}) define completely the evolution of 
$\varepsilon(x,t)$
along the tube, from the onset of the cooling 
procedure till reaching the asymptotic uniform profile
$\varepsilon=+1$ throughout the tube~\cite{foot2}.

For the value $\varepsilon=-1$, the uniform concentration profile
$c_0$ is stable, while for $\varepsilon=+1$ it tends to phase-separate 
(i.e., the initial concentration $|c_0| <\sqrt{1/3}$).  
Therefore, with the advancing cooling front, the $C$-particle concentration 
becomes {\em locally} unstable with respect to phase separation.
As a consequence, a pattern made of alternating low- and high-density
phases of $C$ appears
simultaneously to the propagation of the cooling front along the tube. 
Its properties and characteristics result thus from the
CH equation~(\ref{ch1})
{\em coupled} to the evolution equation~(\ref{ch2}) for $\varepsilon(x,t)$. 
Appropriate boundary conditions (i) guarantee the conservation of
$C$ particles inside
the tube (more precisely, zero-particle fluxes $J_c$ at the edges), 
and (ii) associated to the initial condition, they also ensure the
uniqueness of the solution. The boundary conditions we used in our numerical
discretized procedure amount, in the continuum limit, to
\begin{eqnarray}
&&J_c(x=0\;\mbox{and}\;L,\; t)=0\,,\nonumber\\
&&\frac{\partial^3 c}{\partial x^3}(x=0\;\mbox{and}\;L,\;t)=0\,,
\label{bic1}
\end{eqnarray} 
with 
$J_c(x,t)=\partial\left(\varepsilon c - c^3 +
\partial^2 c/\partial x^2\right)/\partial x$.
Setting $\partial^3 c /\partial x^3=0$ 
means that $c$ at the boundaries relaxes to
$c=\pm\sqrt{\varepsilon}$ determined by the boundary value of $\varepsilon$.
More detailed considerations, including other
types of boundary conditions and
appropriate discretization schemes, are discussed e.g.
in Refs.~\cite{boundary}.

The field $\varepsilon(x,t)$ related to the diffusive temperature profile 
is thus playing the role of a {\em guiding field}. One can think of, however,
other types of fields
$\varepsilon(x,t)$, other boundary and initial conditions
for an experimental setup.
As an example, one can assume that a chemical agent
is diffusing from one reservoir at the $x=0$ end of the tube,
its concentration
changes the local {\em pH} of the system, and thus may drive the
$C$ particles to phase
separation, etc. Accordingly, we  call hereafter $\varepsilon$ the
guiding field, and thus shall not restrict ourselves
to the temperature-like interpretation.

\section{Characteristics of the diffusive guiding field $\varepsilon(x,t)$}
\label{gfc}

During its time evolution, the guiding field $\varepsilon(x,t)$ will modify
locally the position of the spinodal line. At a fixed time $t$, the
spinodal density $c_s=\sqrt{\varepsilon(x,t)/3}$ 
will reach the value $|c_0|$ at a given point $x_f=x_f(t)$, 
therefore initiating locally a phase separation.
The point $x_f(t)$ defines 
the {\em position of the instability front}, which is 
thus determined  
by the condition $\varepsilon(x=x_f,t)=3c_0^2$.
Behind the front, which propagates to the right,
the system becomes locally unstable,
and phase separates into a precipitation pattern of
alternate high- and low-density regions of $C$.

\begin{figure}[htb]\vspace{0.7cm}
\psfrag{XXX}{{\large$v_f \; (L/D)$}}
\centering\includegraphics[width=\columnwidth]{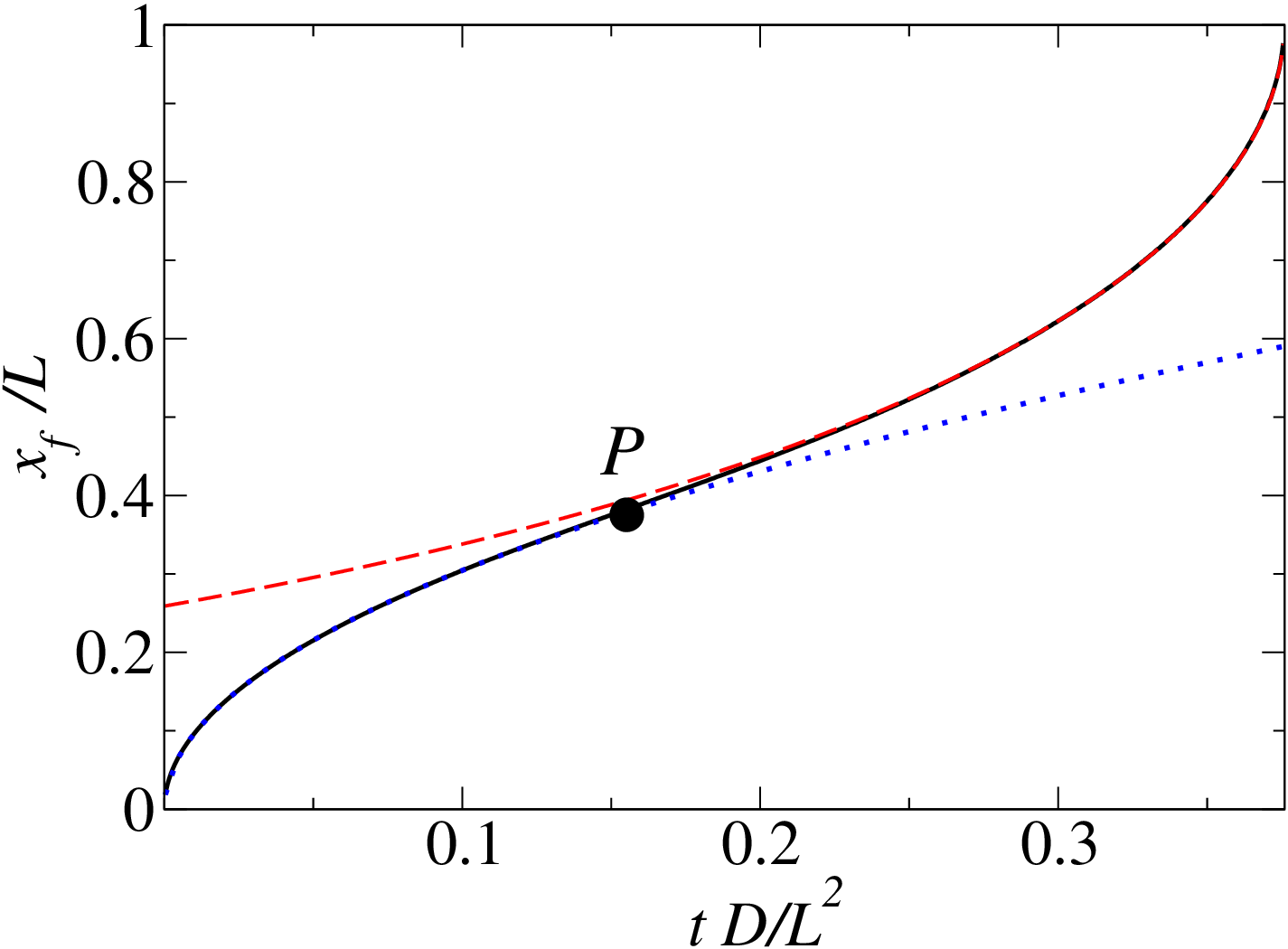}\vspace{1.2cm}\\
\centering\includegraphics[width=\columnwidth]{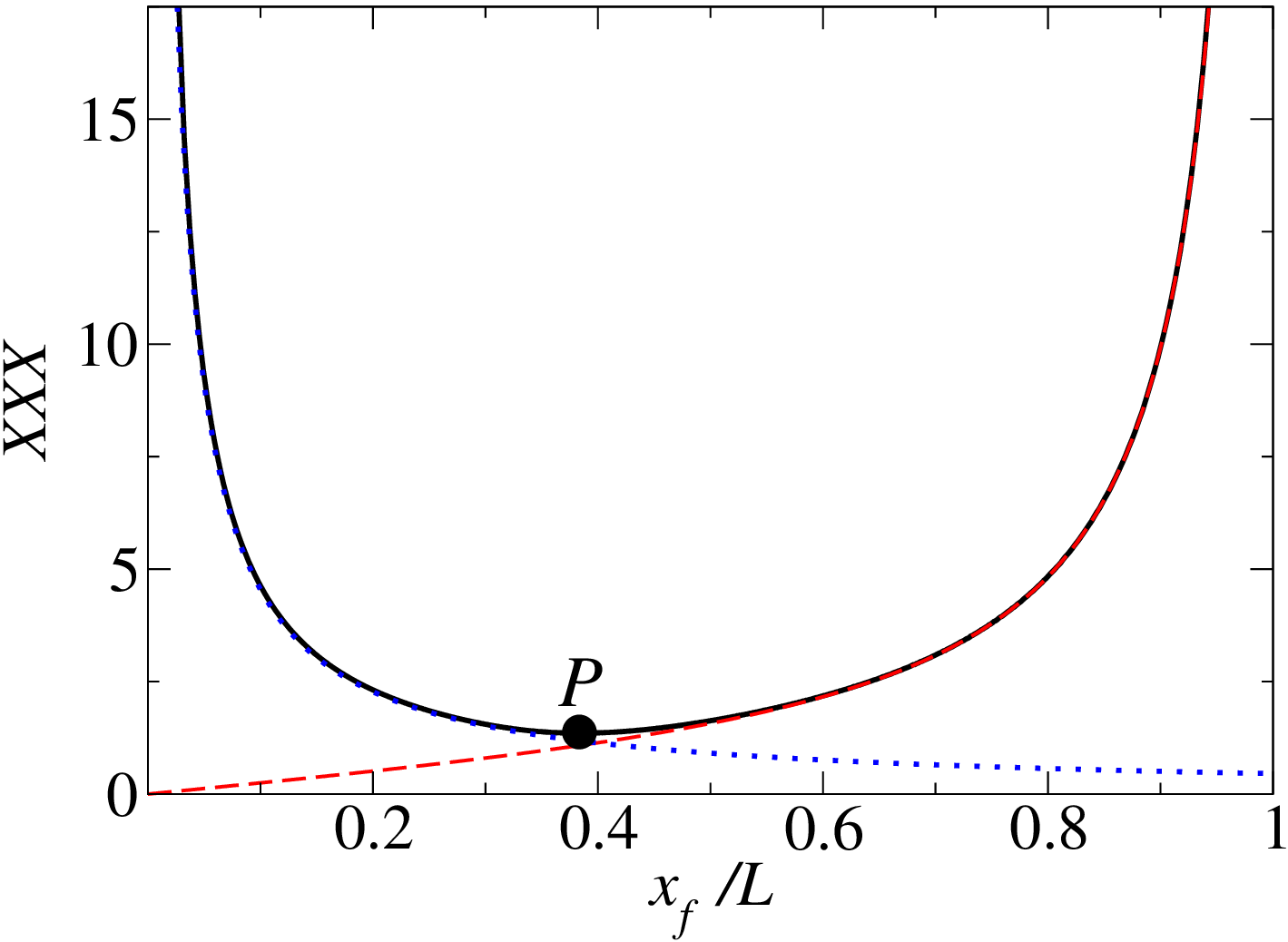}
\caption{Upper panel: Time evolution of the front position
for $c_0=-0.05$ (solid line). The front moves diffusively for small $t$,
$x_f(t) \sim \sqrt{t}$ (dotted line), while the large-time asymptote is
given by Eq.~(\ref{approx-pos}) (dashed line).
Lower panel: The velocity of the instability front $v_f$ as a function of the
front position $x_f$ (solid line). The short-time asymptote $v_f(t) \sim 1/\sqrt{t}$ (dotted line) and large-time behavior given
by Eq.~(\ref{approx-vel}) (dashed line) are also displayed, with
$P$ denoting the crossover point.
The scaling of the spatial, temporal, and velocity variables is described
in the text.}
\label{figure2}
\end{figure}

The diffusion equation~(\ref{ch2}) for $\varepsilon(x,t)$
with the prescribed boundary and initial conditions~(\ref{bic2})
can be solved through a simple Laplace transform method~\cite{crank}.
One obtains for $x_f$ an implicit
equation comprising an infinite sum,
\begin{eqnarray}
&&\sum_{n=0}^{\infty}\frac{(-1)^n}{2n+1}
\exp\left[-\frac{(2n+1)^2\pi^2}{4}\left(\frac{t D}{L^2}\right)\right]
\nonumber\\
&&\times
\cos\left[\frac{(2n+1)\pi}{2}\left(1-\frac{x_f}{L}\right)\right]=
\frac{\pi(1+3c_0^2)}{8}\,.
\label{diff-sum}
\end{eqnarray}
The resulting trajectory of the instability front $x_f(t)$, as well as its
velocity $v_f(t)=dx_f(t)/dt$ for a particular choice of $c_0$ are
represented in Fig.~\ref{figure2}.
Note that when the spatial, temporal, and velocity variables are rescaled,
respectively, by $L$, $L^2/D$, and $D/L$, as indicated on the axis
of these plots, then
the curves for the trajectory and velocity of the instability front
are {\em universal} (for a given value of $c_0$).

As can be seen in Fig.~\ref{figure2},
the front moves diffusively at the beginning, and it accelerates past
a {\em crossover point} $P$ where the acceleration is zero.
The large-time asymptote for the front position can be obtained
by keeping only the leading $n=0$ term in the sum (\ref{diff-sum}),
\begin{equation}
x_f(t)\approx L\left\{1-\frac{2}{\pi}\arccos\left[\frac{\pi(1+3c_0^2)}{8}
\exp\left(\frac{\pi^2Dt}{4L^2}\right)\right]\right\}\,.
\label{approx-pos}
\end{equation}
This approximate expression is valid provided
$L^2/2\pi^2D=t_{min}\lesssim t \lesssim
t_{\rm{max}}=\left(4L^2/\pi^2D\right)\ln\{8/[\pi(1+3c_0^2)]\}$
where $t_{min}$ is the time when the $n=1$ term in the sum (\ref{diff-sum})
becomes negligible with respect to the $n=0$ term, while
$t_{\rm{max}}$ represents a rough
estimate of the time it takes  the instability
front to reach the end ($x=L$) of the tube.
The function given by Eq.~(\ref{approx-pos}) is shown
in the upper panel of Fig.~\ref{figure2}, and one can
see that the asymptote is an excellent approximation past the crossover
point $P$.

The corresponding asymptote for the velocity of the front has the
remarkable property that, when expressed in scaled variables
and in terms of the position of the
front, it becomes independent even of the initial concentration $c_0$,
\begin{equation}
\frac{L}{D}v_f=\frac{\pi}{2}\,\cot{\frac{\pi}{2}\left(1-\frac{x_n}{L}\right)}\,.
\label{approx-vel}
\end{equation}
The above expression is displayed in the lower panel
of Fig.~\ref{figure2} and
one notices again that the approximation is very good past the
crossover point.

\section{Results}
\label{results}

\begin{figure}[h!]\vspace{0.6truecm}
\centering\includegraphics[width=\columnwidth]{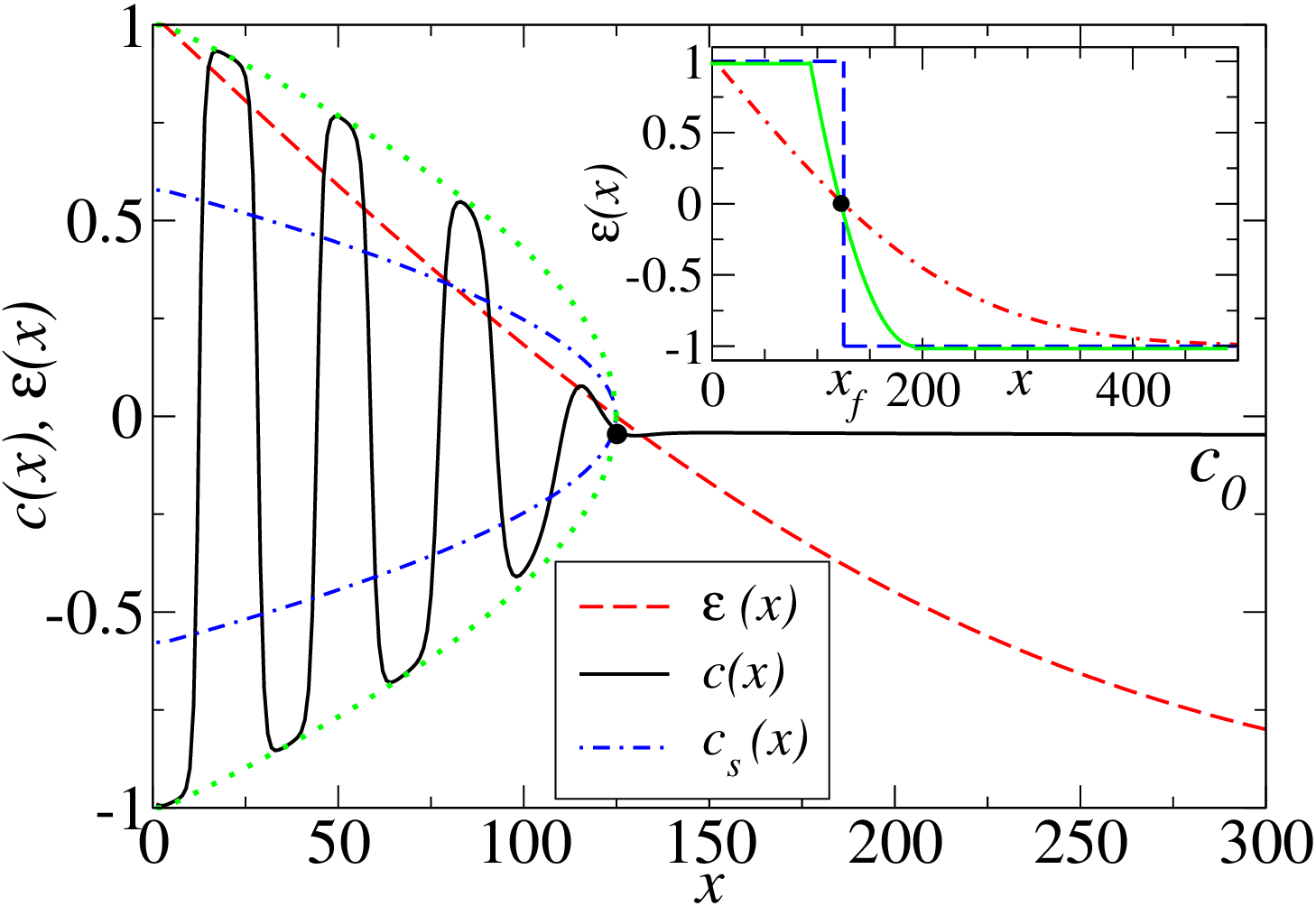}\vspace{0.cm}\\
\caption{Early stage of pattern formation: snapshots of the concentration
field $c(x)$ (continuous line), guiding field $\varepsilon(x)$
(dashed line), 
spinodal lines $\pm
\sqrt{\varepsilon(x)/3}$ (dashed-dotted line) . 
Dotted lines indicate the
local  {\it equilibria} $\pm \sqrt{\varepsilon (x)}$.
The big dot represents the position of the instability front.
The parameters are
$c_0=-0.05$, $L=1000$, $D=4$, and $t=4000$.
The inset shows various
possible profiles of the guiding field, see Sec.~\ref{controling},
namely: the usual diffusive configuration (dashed-dotted line);
a step-like profile (dashed line); and a rigid parabola (continuous line).}
\label{figure0}
\end{figure}

The coupled non-autonomous CH~(\ref{ch1}) and  guiding field~(\ref{ch2}) equations
have been solved numerically for different values of the initial density
$c_0$, diffusion constant $D$, and length $L$ of the tube. Figure~\ref{figure0}
illustrates the early stages of the cooling process, with the
profiles of the concentration $c(x)$,
guiding field $\varepsilon(x)$, and spinodal line
$\pm c_s(x)=\pm \sqrt{\varepsilon(x)/3}$ at a given time $t<t_{\rm{max}}$
(before the instability front reaches the end of the tube).
The concentration field inside the high and low-density
emerging bands relaxes rather rapidly
to the instantaneous, local
{\it equilibrium} values $\pm \sqrt{\varepsilon(x,t)}$, respectively.

The pattern initiated by the instability front evolves afterwards till
reaching a {\it stationary} profile, made of alternate regions of
$c=\pm 1$ and rather sharp interfaces between them, throughout the whole tube. 
Strictly speaking,
this {\it stationary} profile is still evolving through coarsening and
band coalescence, as predicted e.g. in~\cite{langer}. However
(except eventually for some very closely-spaced bands, 
see below the comments on the {\it plug}), its characteristic
evolution time is usually well-beyond any reasonable experimental
time~\cite{simon}; from a practical point of view one can therefore
safely assume its stationarity.

\begin{figure}[h!]\vspace{0.6cm}
\hspace{-0.7cm}\includegraphics[width=\columnwidth]{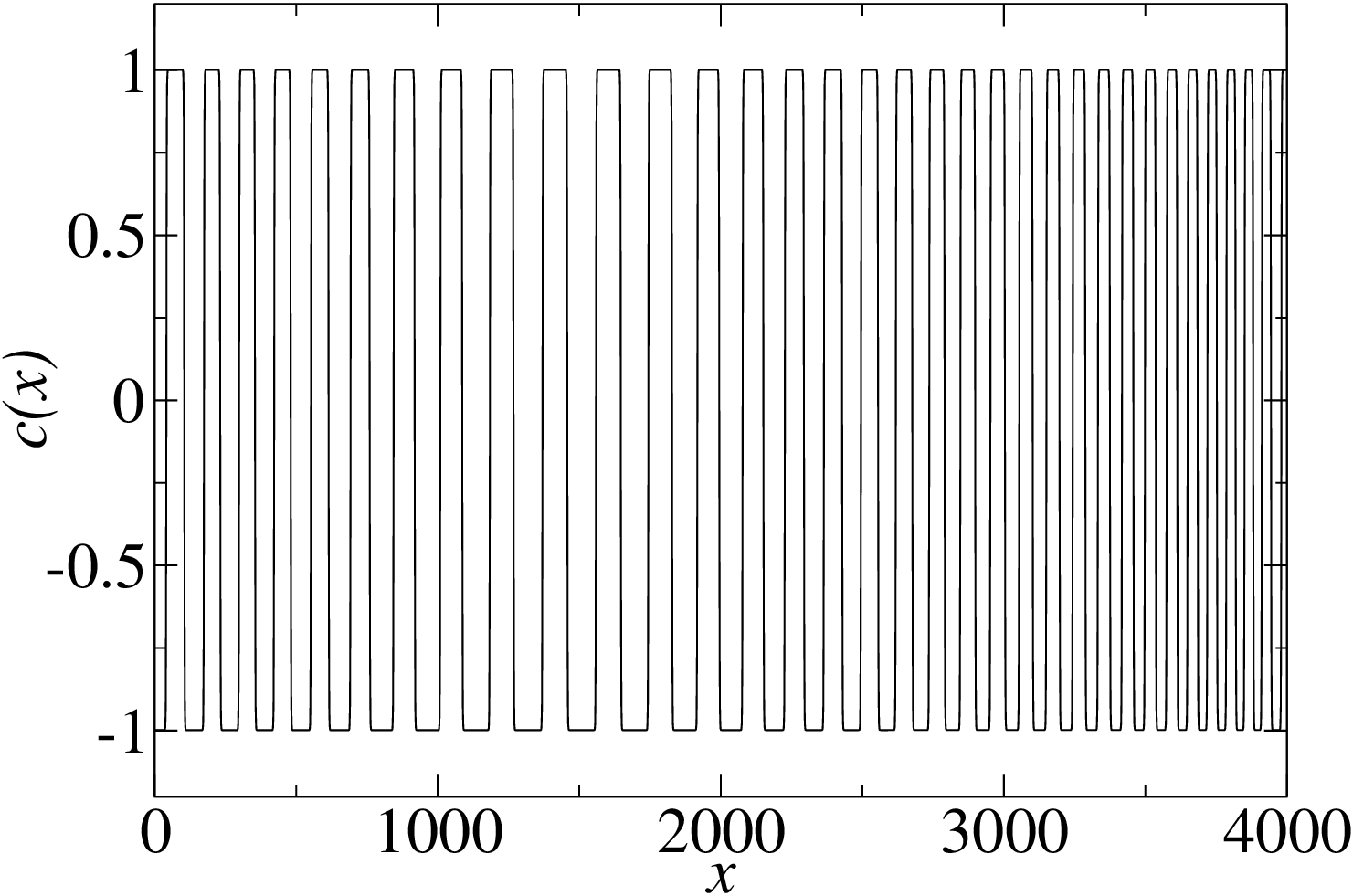}\vspace{1.2cm}\\
\hspace{-0.7cm}\includegraphics[width=\columnwidth]{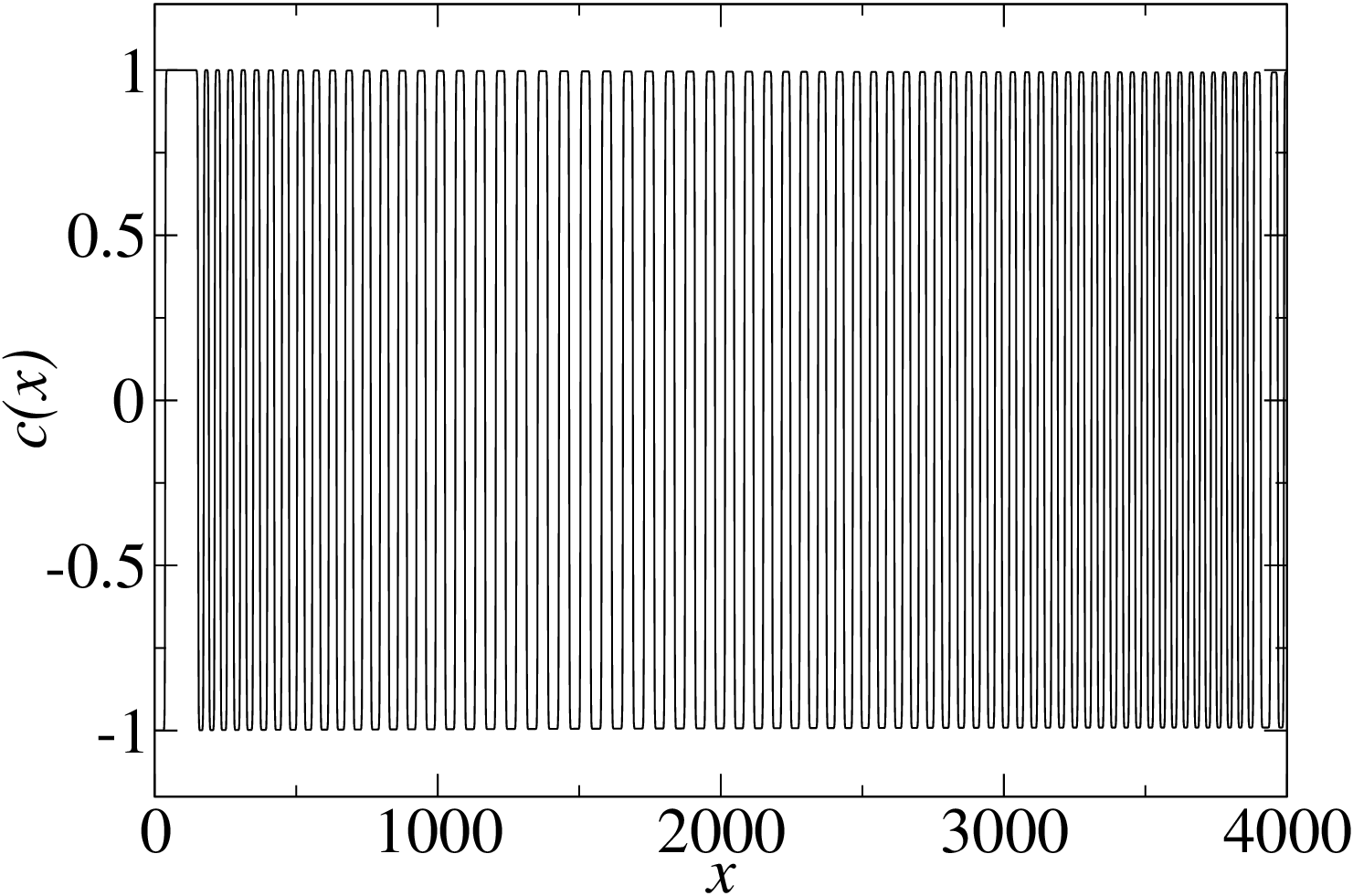}\vspace{1.2cm}\\
\hspace{-0.7cm}\includegraphics[width=\columnwidth]{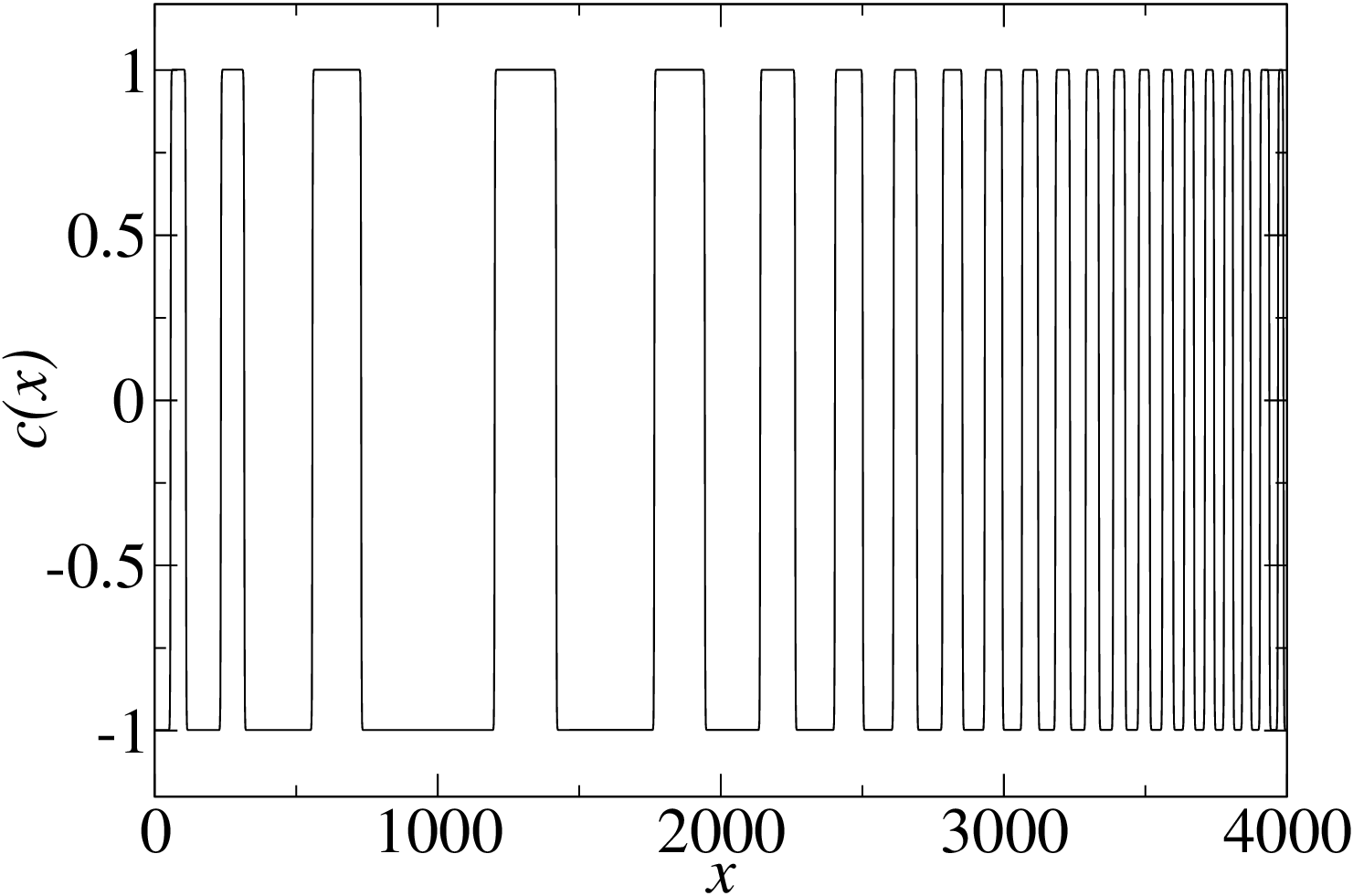}
\caption{The numerical solution $c(x)$ of the non-autonomous Cahn-Hilliard
equation~(\ref{ch1}) in the long-time limit $t \gg t_{\rm{max}}$, and
for system size $L=4000$. Upper panel:
$c_0=-0.05$ and $D=1$; middle panel:
$c_0=-0.05$ and $D=8$; and lower panel:
$c_0=-0.2$ and $D=1$.}
\label{figure1}
\end{figure}

Three typical stationary
patterns of the $C$-particle concentration field
are represented in Fig.~\ref{figure1}.

Before going into a more detailed analysis, let us enumerate some 
general qualitative features of the emerging  patterns:\\
(i) The total number of bands increases as $D$ increases,
for fixed $c_0$ and $L$.\\
(ii) For fixed $D$ and $L$, however, the number of bands decreases
with  increasing $|c_0|$ (approaching the spinodal).\\
(iii) The first part of the pattern displays regular banding
(i.e. increasing distance between consecutive bands),
while a second part displays {\em inverse banding}.
It is important to note that the transition from one
type of pattern to the other is related to the change
in the behavior of the velocity of the
guiding field, namely from the initial diffusive-like motion,
to the later-time accelerated one, see Fig.~\ref{figure2}.\\
(iv) In some situations, the pattern contains an initial
{\it plug}, i.e. a rather wide initial region of constant
concentration, see e.g. the second panel of  Fig.~\ref{figure1}.
This effect has already been encountered
in the usual Liesegang-pattern
formation~\cite{racz99,michel00}. 
The plug may sometimes result from the 
coalescence, on a time scale of the order ${\cal O}(t_{\rm{max}})$, 
of a certain number of very 
closely-spaced bands~\cite{simon}. 
A plug can also form
at the end of the pattern, where the bands can 
be again close enough to each other. 
Contrary to the standard Liesegang pattern
whose spatial extension is only limited by the length of the tube,
in our case the length of the patterned region can thus be {\em limited}
by this band-coalescence effect.

\begin{figure}[htb]\vspace{0.7cm}
\centering\includegraphics[width=\columnwidth]{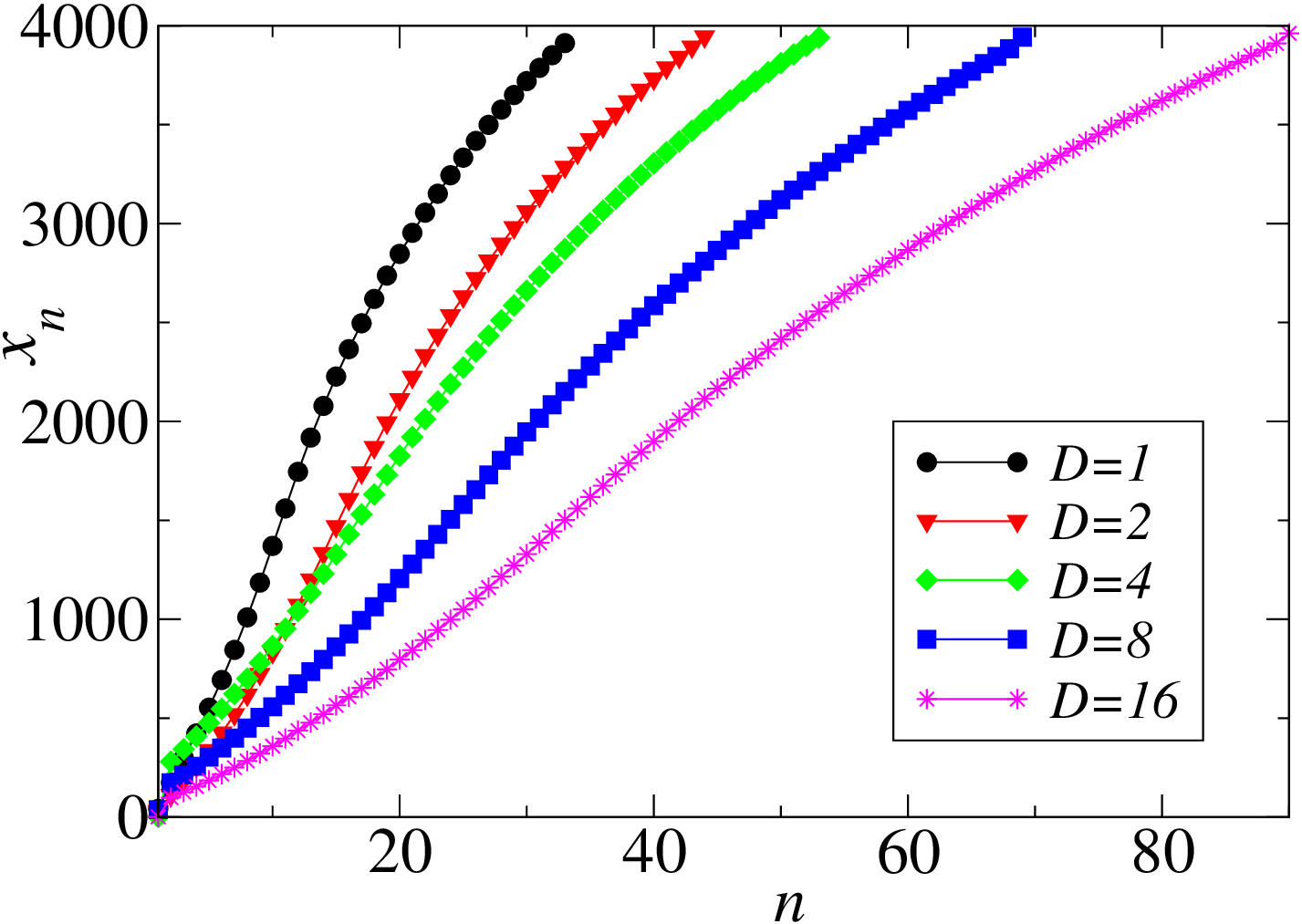}
\caption{The position $x_n$ of the $n$-th band with respect to its 
label $n$ 
for $c_0=-0.05$, $L=4000$, and different values of $D$.}
\label{figure4}
\end{figure}

Let us consider now the characteristics of the patterns
from a more quantitative perspective. 
Figure~\ref{figure4} shows a plot of the band positions $x_n$ 
(which are taken, conventionally, to be the points where $c=0$, 
with an ascendent slope, $dc(x_n)/dx>0$, and are  
enumerated in the order of their appearance, starting
from the $x=0$ end of the tube)
as a function of $n$; different values of the diffusion constant 
$D$ were considered, for fixed $c_0=-0.05$ and $L=4000$.
The presence of a large initial plug 
may have some important
effects on the $n$-dependence of $x_n$ for small $n$ values. 
Accordingly, a simple and experimentally measurable functional
expression is only expected for 
sufficiently large values of $n$,
precisely as in the case of the usual Liesegang patterns.

For the initial, {\it regular-banding} part of the pattern,
if enough bands present, one can
fit the positions of the bands reasonably well with a geometric series,
$x_n \sim \exp(n\tilde{p})$, 
as for a standard  Liesegang pattern~\cite{racz99,michel00}.
This is obviously related to the
initial diffusive-like motion of the instability front,
that does not differ qualitatively from the motion
of the reaction front in the usual Liesegang configuration~\cite{antal99}.
However, a power-law fitting cannot be excluded
either, and further detailed work meant to clarify this point
is in progress and will be published elsewhere. 

For the second, {\it inverse-banding}
part of the pattern, the positions of the bands
for large $n$-s can  be  fitted equally well as $x_n \sim \ln n$ 
or with a power law  
$x_n \sim n^{\beta}$, where the 
exponent $\beta \approx 0.2$--$0.3$   
is practically independent of $D$. 
Since the corresponding distance $\lambda_n=x_{n+1}-x_n$
between consecutive bands behaves 
like $\lambda_n\propto n^{\beta -1}$, the
inequality $\beta <1$  ensures precisely the inverse-banding
character of the pattern.

\begin{figure}[htb]\vspace{0.3cm}
\centering\includegraphics[width=\columnwidth]{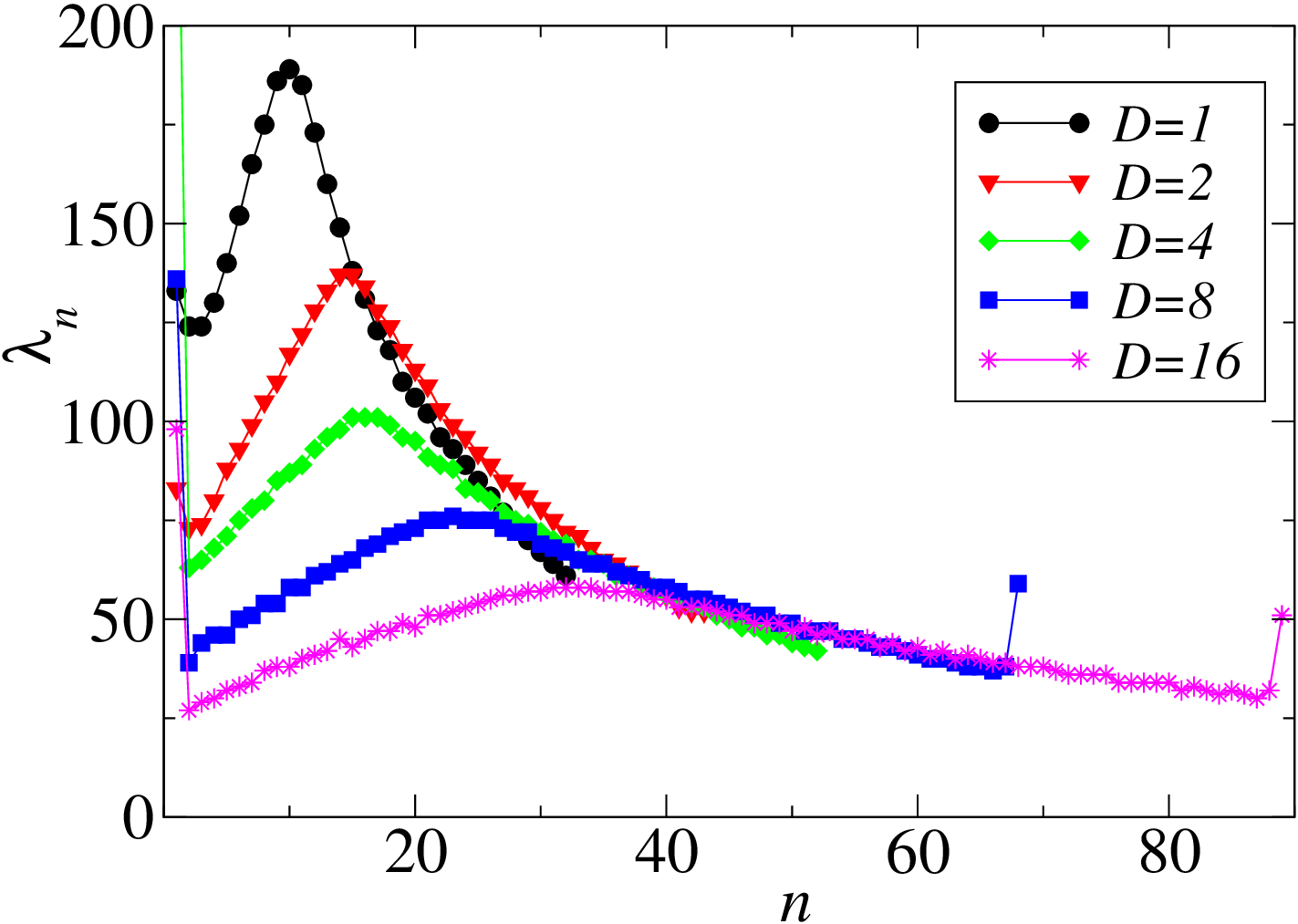}
\caption{The distance $\lambda_n=x_{n+1}-x_n$ 
between two successive bands as a function of $n$,
for $c_0=-0.05$, $L=4000$, and different values of $D$. One can notice
outlier points at the beginning and end of the lines. They correspond to
the initial and final plug regions.}
\label{figure5}
\end{figure}

Figure~\ref{figure5} displays $\lambda_n=x_{n+1}-x_n$
as a function of $n$, for the same
parameter values as in Fig.~\ref{figure4}. One notices clearly the 
initial region of regular banding (with the eventual spurious
initial plug) and the inverse banding, with
the final plug. 
The tails of these plots for large-$n$ values do not allow to discriminate further
between the two above-suggested  fittings of the relation  between $x_n$ and $n$
in the inverse-banding  region.

\begin{figure}[htb]\vspace{0.5cm}
\psfrag{XXX}{{\large$w_n$}}
\centering\includegraphics[width=\columnwidth]{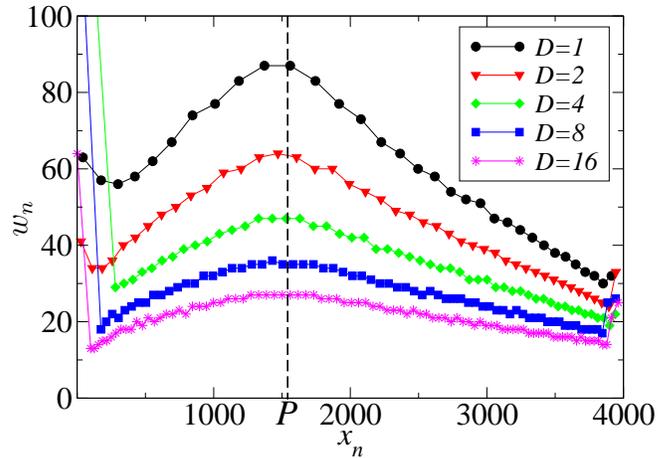}
\caption{The width $w_n$ of the $n$-th band as a function
of the band position $x_n$,
for $c_0=-0.05$, $L=4000$, and different values of $D$.
The vertical dashed line indicates the position
of the crossover point $P$ (see Sec.~\ref{gfc}). 
The jumps at the endpoints of the lines are due to the plugs.}
\label{figure6}
\end{figure}

Finally, in Fig.~\ref{figure6} we plot the width $w_n$
of the $n$-th high-density band as a function of $n$.
It is remarkable that, except for a crossover region between 
direct and inverse banding, one can fit throughout, with a
good approximation:
\begin{equation}
w_n \approx W\,x_n+U\,.
\end{equation}
For the regular-banding region $W>0$, and one can easily
justify this result simply by using mass-conservation arguments for the
$C$-particles, as well as the geometric progression of band positions,
see~\cite{racz99,michel00}.
However, for the inverse banding region $W<0$
and the approximate nature of this 
relationhip is 
related to the fact that the band
positions are not well-fitted by a geometric
series in this case.
Note that on this figure one can clearly see that, as already stated above, 
the transition from
regular to inverse banding is marked by the crossover point $P$
of the motion of the instability front, see Sec.~\ref{gfc}.

\section{Theoretical arguments}
\label{interpretation}

Our goal is to devise a simple theoretical approach 
able to explain the characteristics of the
patterns observed in the numerical simulation of the non-autonomous
CH equation~(\ref{ch1}), and to be used further on
for predictive purposes. 
A  basic element of our approach is the numerical finding that
the characteristics of the patterns are directly related
to the motion of the instability front. In particular, the
{\it local wavelength} of the pattern $\lambda_n=x_{n+1}-x_n$
(see Figs.~\ref{figure5}-\ref{figure6}) is
related to the velocity $v_f=v_f(x_f)$ of the front
(see the second panel of Fig.~\ref{figure2}), as discussed below.

Our approach is based on several assumptions, 
the validity of which is
verified a-posteriori by comparison of the theoretical
findings with  the results of the numerical simulations. 
Our first hypothesis is that the
guiding field moves faster than the diffusing $C$ particles;
therefore, the phase-separation does not take place
ahead the instability front, but only behind it. Note, however, that
the velocity of the front should not be too high either,
since otherwise our second hypothesis about the quasi-stationarity
may not be fulfilled.
The meaning of the second hypothesis is that although the local value of the
spinodal concentration $c_s(x,t)=\sqrt{\varepsilon(x,t)/3}$
evolves in time in the wake of the instability front,
this evolution can be assumed to be slow enough, so that
the local instability boundaries associated with the
spinodal curve are in a quasi-stationary state.
We assume therefore that the onset of the phase-separation instability
is {\it pulled} by the motion of the guiding front,
and consequently we can use
the standard results of the {\em pulled-front}
theory~\cite{vansar1,vansar2,vansar3}
to establish the characteristics of the emerging pattern.

Let us recall here the main results of
the standard theory. Consider an autonomous
CH equation~(\ref{ch1}) with
$\varepsilon={\rm constant}$ throughout the system,
and a uniform unstable concentration
$c_0$, $|c_0|<c_s=\sqrt{\varepsilon/3}$. 
A sharply-localized perturbation of this state
will then evolve into an instability front,
with a well-defined velocity, leading  to 
phase-separation behind it and to the
appearance of a pattern of well-defined wavelength.
Using linear stability analysis arguments, one can easily compute
both the wavelength $\lambda^*$
of the most unstable mode and the {\em asymptotic} 
velocity $v^*$ of the instability front as a function of the distance
between the initial concentration and the spinodal value.
Namely,
\begin{eqnarray}
\lambda^* &=& \frac{16\pi \sqrt{2} (\sqrt{7}+2)}
{3(\sqrt{7}+3)^{3/2}}\,a^{-1/2}\,,   
\\
v^* &=& \frac{2(\sqrt{7}+2)}{3(\sqrt{7}+1)^{1/2}}\,a^{3/2}\,,
\label{asvel}
\label{autonomous}
\end{eqnarray}
where $a \equiv 3(c_s^2-c_0^2)=\varepsilon-3c_0^2$. 
Except for the cases when one has
band-coalescence (coarsening), this wavelength provides
the wavelength of the {\em asymptotic} emerging pattern.
By eliminating the parameter $a$ between these two expressions,
one obtains a direct relationship between
the asymptotic wavelength of the pattern and the asymptotic velocity of
the instability front,
\begin{equation}
\lambda^* =\frac{9.642}{(v^*)^{1/3}}\,.
\label{dr}
\end{equation}
Note however that the relaxation of the system to this
asymptotic state goes rather slowly, like $(1/t)$, both for the 
wavelength of the 
pattern and for the velocity of the instability front. Moreover,
the transient effects tend to increase the wavelength of the pattern
above its asymptotic value $\lambda^*$, see~\cite{vansar1,vansar2,vansar3} for
further details. \\

Using the above results of the pulled-front theory,
we make now the Ansatz that
the relationship~(\ref{dr}) remains valid for our non-autonomous
CH equation. More precisely,
we assume that the local wavelength of the pattern is determined
by the instantaneous/local velocity of the instability front as
\begin{equation}
\lambda_n \approx \frac{9.642}{\left[v_f(x_n)\right]^{1/3}}\,.
\label{FORMULA}
\end{equation}
The physical picture underlying the above assumption is the following.
The instantaneous
pulled-front velocity $v^*=v_f(t)$ {\it dictates}, see Eq.~(\ref{asvel}),
an instantanous value of the parameter $a$, let us call it $a_f(t)$.
This means that the local concentration in the vicinity of the
quasi-stationary instability front adjusts rapidly to
the value $c_f(t)$ corresponding to the parameter $a_f$, namely
$a_f(t)=\varepsilon(x=x_f(t),t)-3c_f^2(t)$.

\begin{figure}[htb]\vspace{0.7cm}
\centering\includegraphics[width=\columnwidth]{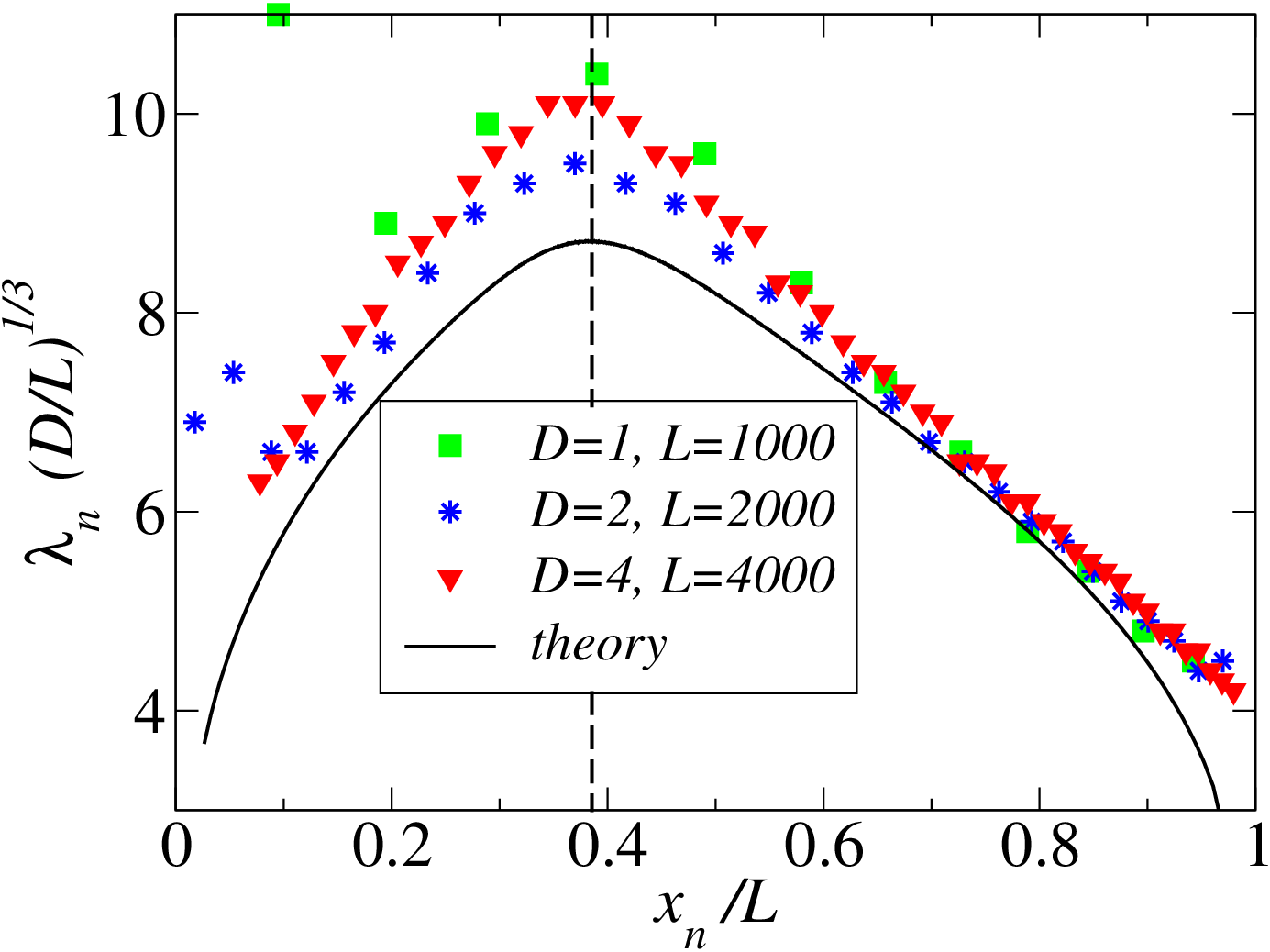}\vspace{1.2cm}\\
\centering\includegraphics[width=\columnwidth]{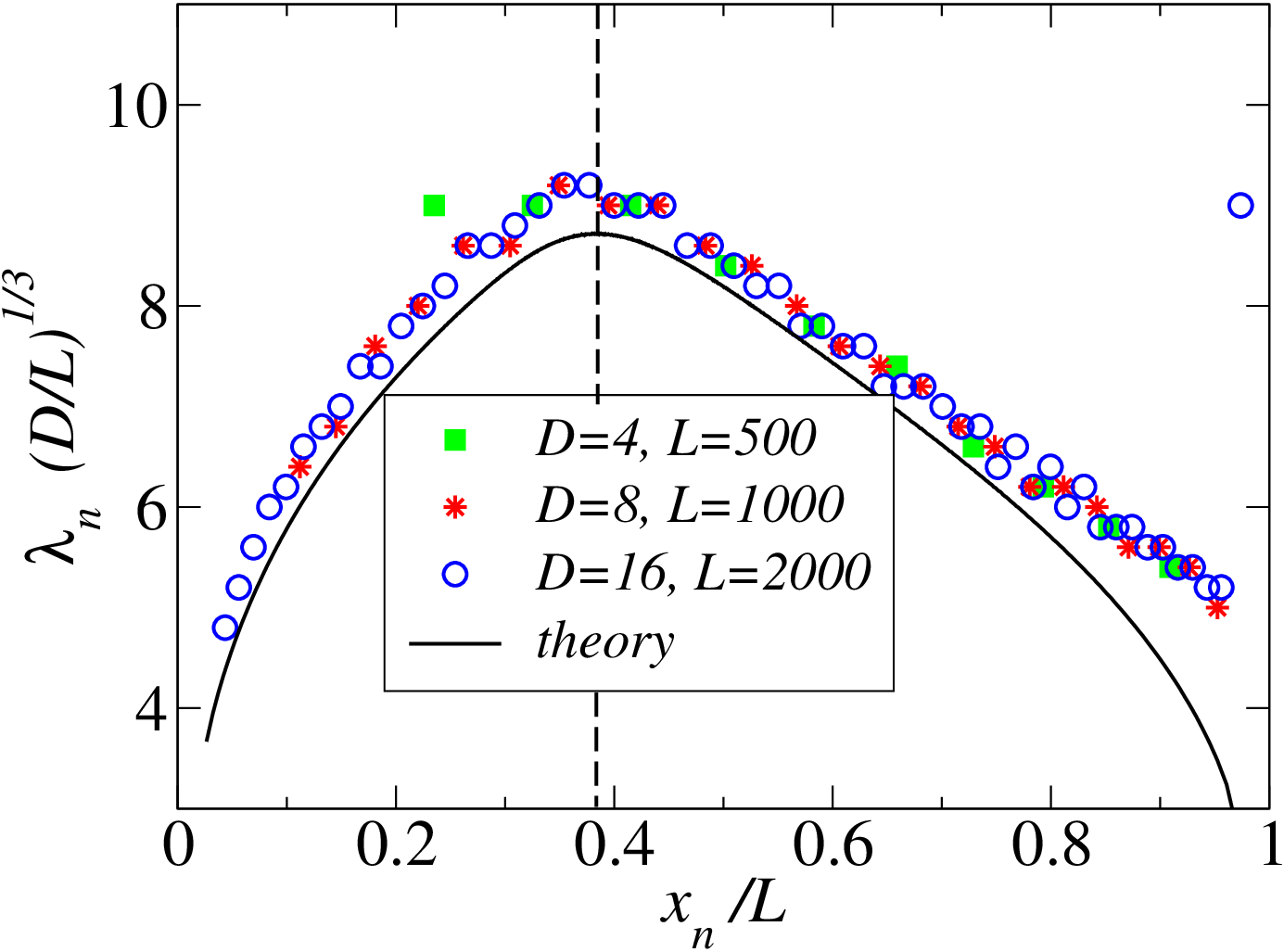}
\caption{Comparison of the local wavelengths of the pattern
as determined from numerical simulations (symbols)
with the theoretical results based on Eq.~(\ref{FORMULA})
(continuous line).
The wavelengths $\lambda_n$ and the band positions $x_n$,
are rescaled as discussed in the text.
Upper panel: $c_0=-0.05$ and $L/D=1000$. Lower panel: $c_0=-0.05$ and
$L/D=125$. The vertical dashed lines indicate the position
of the crossover point $P$ where the instability front
has a minimal velocity
(see Sec.~\ref{gfc}). The outlier endpoints
are due to the plugs.}
\label{figure7}
\end{figure}

The comparison of the theoretical findings based on
the above Ansatz with the results of the numerical simulations
is displayed in Fig.~\ref{figure7},
where the local wavelength of the pattern
$\lambda_n$ is plotted versus $x_n$
for different values of $D$ and $L$. This figure provides
a double-check of the Ansatz.  Namely,\\
(i) If Eq.~(\ref{FORMULA}) is valid, then, since $v_f(x_n)$ is universal under
appropriate scaling of space, time, and velocities (according to
Sec.~\ref{gfc}), then the plots from the numerical results should merge
when applying the rescaling
$\lambda_n \rightarrow \lambda_n  (D/L)^{1/3}$
and $x_n \rightarrow x_n/L$. This is, indeed, the case, as illustrated by both
panels of Fig.~\ref{figure7}.\\
(ii) All the rescaled plots should fit the theoretical
formula~(\ref{FORMULA}) shown by solid lines in Fig.~\ref{figure7}.

The agreement between our simple theoretical predictions and
the results of the simulations is surprisingly good. 
The only  exceptions are a few {\it outlier} points
corresponding, respectively,  to  the early band formation (initial plug) and to the 
last bands close to the boundary (final plug). 
There is also a systematic initial and final mismatch, that
may be  due to the high acceleration of the instability
front at the very beginning and the very end of its motion along the tube
(see Fig.~\ref{figure2}), and thus to the breaking of the quasi-stationarity
hypothesis that lies at the basis of our Ansatz. Another origin of
discrepancy can be the dynamics of the guiding field profile,
that may, under some circumstances, fail the quasi-stationarity hypothesis.

We note that the  agreement between numerics and theory is better for large
values of the velocity of the instability front, i.e., for smaller values 
of $L/D$, as shown in the upper panel of Fig.~\ref{figure7} as compared to the lower panel. This is probably related to a better, respectively worse
adequacy of the basic hypothesis of a
fast-moving front as compared to the diffusion 
of the $C$ particles. Finally, the fact that the numerical wavelengths are
systematically larger that the theoretically-estimated ones may be 
the combined effect of slow relaxation to the asymptotic state and the
quasi-stationary nature of our configuration during the onset of the pattern
(see the comments above on the effects of transients on the wavelength, in the
autonomous case). 

\section{Pattern control}
\label{controling}

We address now the problem of {\em controllability}
of the characteristics of the emerging pattern. It is
obvious from the above results that both the qualitative
(i.e., regular or inverse banding)
and quantitative features (like total length of the pattern,
pattern local wavelength, width of bands, etc.) 
can be controlled in the described configuration through 
an appropriate choice of the parameters  $L$,  $c_0$, 
and (to a less extent, as more difficult to manipulate) $D$.
Moreover, these results can be described  
theoretically in the frame of the {\it pulled-front}
approximation thus providing a method for estimating the
parameters of the patterns.
However, this method of control, although very simple, 
is somewhat rigid, 
since the above-mentioned control parameters
cannot be changed during the process, while, ideally,
one requires an easily tuned, flexible, external 
tool of control. One can then think about moving the
tube with the gel (or maybe a thin film of gel)
in a prescribed temperature profile, with a velocity 
that can be changed at any moment according to the needs. 
One achieves therefore a guiding field $\varepsilon(x,t)$ that can 
be externally tuned at any moment and point. 

For example, the simplest configuration one can imagine 
is an abrupt, {\em step-like} temperature profile that 
moves with velocity $v_f$, such that 
$\varepsilon(x,t)=-1+2 \Theta(x-x_f(t))$ 
($0\leqslant x \leqslant L$), where $\Theta(\dot)$ 
designates the Heaviside step function
and $x_f(t)$ is the instantaneous position of the step. 
If the motion is uniform
$v_f=\rm{const.}$, then one obtains {\em equidistant banding}. If
the motion of the step is accelerated or decelerated, 
then the pattern presents 
{\em inverse-banding}, respectively {\em regular banding}, 
with characteristics
that depend on the details of $v_f=v_f(t)$.

Another simple option is to propagate a smooth, given
temperature profile along the tube, such that
$\varepsilon(x,t)=F(x-x_f(t))$.
Now, the characteristics of the
emerging pattern do not depend only on
the velocity $v_f(t)$ of the propagating {\it rigid}
guiding field profile, but also on the shape of this profile.

\begin{figure}[htb]\vspace{0.6cm}
\centering\includegraphics[width=\columnwidth]{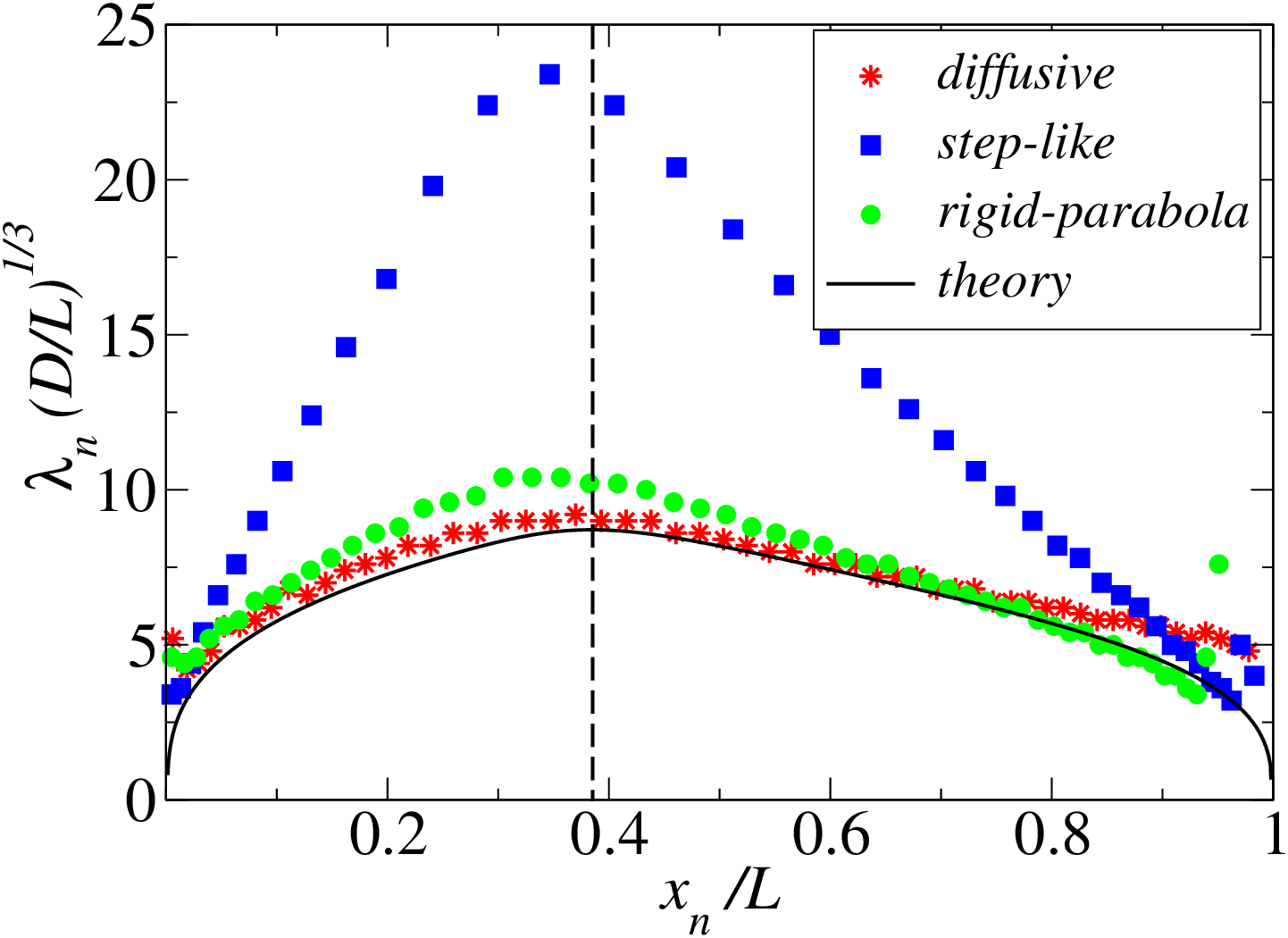}
\caption{Local wavelengths of the pattern $\lambda_n$
versus band positions $x_n$, in appropriate rescaled variables.
Symbols: numerical simulations
for three different profiles of the guiding field 
(see the text).
Continuous line: theoretical calculations based on Eq.~(\ref{FORMULA}).
The parameters are  $c_0=-0.05$, $D=16$, and $L=2000$.
The vertical dashed line indicates the position, along the cylinder's axis,
of the crossover point $P$ where the instability front has a minimal velocity
(see Sec.~\ref{gfc}). The outliers at
the beginning and end of the lines correspond to the plugs.}
\label{figure8}
\end{figure}

In order to illustrate these points, we considered, for
comparison, the emerging pattern in three situations (also illustrated
in Fig.~\ref{figure0}),
namely for:\\
(i) The diffusive guiding-field profile, as discussed
in the previous Sections, for a given set of parameters $D$, $L$, and $c_0$.
Recall that the instability front moves with a velocity 
$v_f(t)$ described in Sec.~\ref{gfc};\\
(ii) A step-like profile of the guiding field that moves
with the same velocity $v_f(t)$;\\
(iii) Finally, a parabolic profile of guiding field,
$\varepsilon(x,t)=\left[-1+{\cal A}(x_f(t)-x-x_0)^2
\Theta(x_f(t)-x-x_0)\right]$.
One has $\varepsilon=-1$ for $x=x_f+x_0$ and the parameter
${\cal A}$ is determined such that for $x=x_f$ 
one has $\varepsilon=1-3c_0^2$. 
As before, this rigid profile 
moves with the same velocity $v_f(t)$.\\
The results of the numerical simulation are represented in Fig.~\ref{figure8},
together with the theoretical result based on our Ansatz.
One can  notice that:\\
(i) The pattern can be effectively controlled by the proposed methods.
The effects are qualitatively the same as for our usual configuration,
but  these new methods allow for more flexible control.\\ 
(ii) The pattern obtained for the parabolic-like profile (with 
$x_0=70$) is closer to the pattern obtained for our usual configuration, as
well as to the theoretical predictions based on the pulled-front
approximation;
the pattern obtained for the step-like profile is much different.
This  convincingly illustrates the importance of the quasi-stationarity
hypothesis for the pulled-front theory. Indeed, this basic
ingredient is a good approximation 
both for our usual configuration 
and for the parabolic profile, but it is definitely absent
in the case of the step-like profile, for which the associated 
abrupt jump in the local value of $\varepsilon$ forbids any possibility
of quasi-stationarity.

\section{Conclusion}
\label{conclusions}

We have discussed the problem of how to control precipitation
patterns by bringing a system described by the Cahn-Hilliard
equation into an unstable state using a prescribed
guiding field. It was shown that
simple, physically realizable fields,
such as a temperature field generated by a
temperature jump at the boundary, is sufficient to generate
rather complex precipitation patterns even in one dimension.
The spacing characteristics of the
patterns were determined numerically for the case of a
diffusive guiding field, and we developed a quantitative theory
for explaining the simulation results. The theory
is based on relating the velocity of the instability front generated
by the guiding field to the natural, pulled-front velocity
of the phase-separation process which, in turn, controls the
lengthscale of the pattern left in the wake of the moving front.

From a theoretical point of view, our results suggest that the
inverse-banding phenomena observed in some Liesegang experiments
may have an explanation in terms of a diffusive guiding field.
This guiding field is perhaps not a temperature field, but may be
generated by the diffusion of some chemical species which do
not take part in the reactions and the precipitation but may change e.g.
the local {\em pH} value and thus influences the precipitation thresholds.

As far as the technological applications are concerned, it appears
that the problem of microfabrication of bulk
structures by chemical reactions and
precipitation~\cite{grzy3,grzy2,grzy1,optical,tsapatis,giraldo}
is just in the first stages of its development. The usefulness
of this field will be decided on the possibility of creating
flexible ways to ensure controllability.
Our results suggest experimentally feasible solutions for the
control of a particular precipitation process (formation of Liesegang
bands).
Clearly, further studies are necessary  to develop new methods of control
and to sort out the question of controllability in more complex cases.

\acknowledgments
This research has been partly supported by the Swiss National
Science Foundation and
by the Hungarian Academy of Sciences (Grant No.\ OTKA T043734). Financial
support is also acknowledged to
the Program for Evolutionary Dynamics
at Harvard University by Jeffrey Epstein and NIH grant R01GM078986 (TA).


\begin{thebibliography}{}
\bibitem{Note-on-C} {Note that the reactional scheme can be actually rather complex. In particular, it can involve the possible existence of an intermediate compound, see e.g.\protect \nobreakspace  {}\cite {antal98,droz99} for a more detailed discussion of these aspects and of related theoretical models.}
\bibitem{foot2} The shift and rescale
of the concentration were done such that for the asymptotic state
$\varepsilon=+1 $ the stable uniform concentrations correspond to $c=\pm 1$.
This can be achieved as
$$c=\displaystyle\frac{c\;{\rm(unscaled)}\;-(c_{\rm h}+c_{\rm l})/2}
{(c_{\rm h}-c_{\rm l})/2}\;,$$
where $c_{\rm {h,l}}$ denote, respectiely, the unscaled high and low-density equilibrium
phases concentrations for $T=T_0$ ($\varepsilon=+1$).
\end{thebibliography}
\end{document}